\documentstyle[preprint,aps]{revtex}
\begin{document}
\preprint{KNU-T0112}
\def\a{\alpha}
\def\b{\beta}
\def\e{\epsilon}
\def\g{\gamma}
\def\p{\partial}
\def\m{\mu}
\def\n{\nu}
\def\t{\tau}
\def\s{\sigma}
\def\half{\frac{1}{2}}
\def\hatt{{\hat t}}
\def\hatx{{\hat x}}
\def\hatp{{\hat p}}
\def\hatX{{\hat X}}
\def\hatY{{\hat Y}}
\def\hatP{{\hat P}}
\def\hatth{{\hat \theta}}
\def\hatta{{\hat \tau}}
\def\hatrh{{\hat \rho}}
\def\hatva{{\hat \varphi}}
\def\p{\partial}
\def\nn{\nonumber}
\def\barx{{\bar x}}
\def\barY{{\bar Y}}
\def\cb{{\cal B}}
\def\2pap{2\pi\alpha^\prime}
\def\wideA{\widehat{A}}
\def\wideF{\widehat{F}}
\def\beq{\begin{eqnarray}}
\def\eeq{\end{eqnarray}}
\def\2pap{2\pi\a^\prime}

\title{Tachyon Condensation and Open String Field theory}
\author{{Taejin Lee} 
\thanks{E-mail: taejin@cc.kangwon.ac.kr}}
\address{{\it 
Department of Physics, Kangwon National University, 
Chuncheon 200-701, Korea}}

\date{\today}
\maketitle
\begin{abstract}
We perform canonical quantization of the open string on a 
unstable D-brane in the background of the tachyon condensation. 
Evaluating the Polyakov path-integral on a strip, we obtain the 
field theoretical propagator in the open string theory. 
As the condensation occurs the string field theory is continuously 
deformed. At the infrared fixed point of the condensation, the open 
string field on the unstable D-brane transmutes to that on the lower 
dimensional D-brane with the correct D-brane tension.  
\end{abstract}

%04.60.Ds Canonical quantization
%11.25.-w Theory of fundamental strings
%11.25.Sq Nonperturbative techniques; string field theory
%11.25.H Algebraic methods in string theory
%11.27 Strings extended classical solutions, 

\narrowtext

\section{Introduction}

The tachyon condensation is a noble phenomenon in string theory, 
which determines the ultimate fates of the unstable D-branes and 
the $D$-${\bar D}$-brane pairs. The unstable systems in string 
theory are expected to reduce to the stable lower dimensional 
D-brane systems or disappear into the vacuum, leaving only the closed 
string spectrum behind. Since the celebrated Sen's conjecture
\cite{sen} on the tachyon condensation many important aspects of 
this noble phenomena have been explored by numerous authors. Since the 
tachyon condensation is the off-shell phenomenon, the theoretical 
framework to deal with it should be the second quantized string theory. 
The main tools to discuss the tachyon condensation are the 
open string field theory with the level truncation \cite{level} and 
the boundary string field theory \cite{bisft,bsft}. The former one, which is based on the 
Witten's cubic open string field theory \cite{cubic}, has been a useful practical tool to 
describe the decay of the unstable D-branes to the bosonic string vacuum. 
The latter one, which is based on the background independent string field 
theory, has been useful to obtain the effective tachyon potential. 
These two approaches are considered to be complementary to each 
other. 

In a recent paper \cite{tlee} we discuss the tachyon 
condensation in a single D-brane, using the boundary 
state formulation \cite{callan,bsf}, which is closely related 
to the latter one. As we point out, the boundary 
state formulation contains the boundary string field theory, 
since the normalization factor of the boundary state corresponds to 
the disk partition function, which is the main ingredient of the latter 
approach. Moreover, it provides an explicit form of the quantum state 
of the system in terms of the closed string wavefunction. 
Thus, we may find a direct connection between the boundary state 
formulation and the former approach based on the string field 
theory if we appropriately utilize the open-closed string duality. 
It suggests that the succinct boundary state formulation of the
tachyon condensation may be transcribed into the open string field 
theory. The purpose of this paper is to construct the open string field 
theory in the background of the tachyon condensation and to show that 
the descent relations among the D-branes is also well described in 
the framework of the open string field theory. To this end we 
perform canonical quantization \cite{tlee2} of the open string on a 
unstable D-brane in the background of the tachyon condensation.
Then we evaluate the Polyakov string path-integral on a strip
to obtain the field theoretical propagator of the open string theory
in the background of the tachyon condensation. 
At the infrared fixed point of the condensation, the open 
string field on the unstable D-brane transmutes to that 
on the lower dimensional D-brane with the correction D-brane tension. 

\section{Canonical Quantization}

It is well known that the field theoretical string propagator is 
obtained from the first quantized string theory, by evaluating the 
Polyakov path-integral over a strip, which is the world-sheet of 
the open string in this case. Following the same steps, we will 
construct the field theoretical open string propagator in the 
background of the tachyon condensation. To this end we perform 
canonical quantization of the open string attached on D-brane in the 
tachyon background. Then integration over the proper time yields 
the string propagator, therefore the kinetic part of the second 
quantized string theory. As we vary the parameter of the tachyon
profile, the field theoretical action for the open string on a 
D-brane is continuously deformed and eventually reduced to that
on a lower dimensional D-brane. For the sake of simplicity we consider 
the bosonic string on a single D-brane. Extension to more general 
cases is straightforward.

The action for the open string in the background of the tachyon 
condensation is given as \cite{com1} 
\beq \label{action}
S = S_M + S_T = - \frac{1}{4\pi \a^\prime} \int_M d\t d\s \sqrt{-h} 
h^{\a\b} g_{\m\n} \p_\a X^\m \p_\b X^\n + \int_{\p M} d\t N T(X)
\eeq
where we consider a simple tachyon profile, 
$T(X) = u_{ij} X^i X^j$. 
Here $N$ is an einbein on the world-line of 
the end points of the open string and its relation to the world-sheet 
metric is given by
\beq
\sqrt{-h} h^{\a\b} = \frac{1}{N} \left(\begin{array}{cc}
  -1 & 0 \\
  0 & N^2 
\end{array} \right).
\eeq
That is, $N$ is the lapse function of the world-sheet metric.
The string action is manifestly invariant under the 
reparametrization.
%where
%\beq
%\frac{d \t}{d \t^\prime} N(\t) = N^\prime (\t^\prime).
%\eeq
We may fix this reparametrization invariance by choosing the 
proper-time gauge, $\frac{d N}{d \t} = 0$, equivalently $N = T$,
constant. Hereafter we confine our discussion to the proper-time 
gauge.
The string propagator is defined as a Polyakov path-integral 
over a strip \cite{tlee3}
\beq
G[X_f; X_i] = \int D[N] D[X] \exp \left(i S_M + iS_T \right) 
\eeq
where the path integral is subject to the boundary condition
$X^\m (\t_f, \s) = X^\m_f (\s)$, $X^\m (\t_i, \s) = X^\m_i (\s)$.

In order to understand the structure of the open string propagator 
on a $D$-brane, let us first consider a flat D-brane where 
$S_T= 0$. Introducing the canonical momenta $P_\m$,
we find that the propagator is written as 
\beq
G[X_f; X_i] = \int^\infty_0 d T \int D[X,P] e^{i \int^T_0 
\left(\int P_\m \dot{X}^\m d\s - H \right)d\t}.
\eeq
If we expand the canonical variables in terms of normal modes
$ X^\m(\s) = \sum_n X^\m_n e^{in\s}$, 
$P_\m(\s) = \sum_n P_{\m n} e^{-in\s}$,
we find that the Hamiltonian is given as 
\beq
H = \frac{1}{2} \sum_n g_{\m\n} \left((\2pap)P^\m_{n} P^\n_{-n} +
\frac{n^2}{(\2pap)} X^\m_n X^\n_{-n} \right). 
\eeq
%The real conditions for the canonical variables read as 
%$X^{\m *}_n = X^\m_{-n}$, $P^*_{\m n} = P_{\m -n}$.
For the open string on a $Dp$-brane in $d$ dimensions, we need to impose 
the Neumann boundary condition for $X^i$ and the Dirichlet 
boundary condition for $X^a$;
$\p_{\s} X^i|_{\p M} = 0$, $X^a|_{\p M} = 0$
where $i = 0, 1, \dots, p$, $a = p+1, \dots, d-1$. 
These boundary conditions result in the following constraints
\begin{mathletters}
\beq
X^i_n - X^i_{-n} &=& 0, \quad P^i_n - P^i_{-n} = 0, \\
X^a_n + X^a_{-n} &=& 0, \quad P^a_n + P^a_{-n} = 0, \\
x^a &=& 0, \quad p^a = 0 \nn
\eeq
\end{mathletters}
for $n = 1, 2, \dots$.
Thus, the canonical variables are written as 
\beq
X^i &=& x^i + \sqrt{2} \sum_{n=1} Y^i_n \cos n\s, \quad
P^i = p^i + \sqrt{2} \sum_{n=1} K^i_n \cos n\s,\nn \\
X^a &=& \sqrt{2} \sum_{n=1} {\bar Y}^a_n \sin n\s, \quad 
P^a =  \sqrt{2} \sum_{n=1} {\bar K}^a_n \sin n\s \nn
\eeq
where $(Y_n, K_n)$ and $(\bar{Y}_n, \bar{K}_n)$ form canonical 
pairs, $Y^\m_n = 1/\sqrt{2} (X^\m_n + X^\m_{-n})$, 
$\bar{Y}^\m_n = i/\sqrt{2} (X^\m_n - X^\m_{-n})$,
$K^\m_n = 1/\sqrt{2} (P^\m_n + P^\m_{-n})$,
$\bar{K}^\m_n = - i/\sqrt{2}(P^\m_n - P^\m_{-n})$.
The procedure given above is equivalent to reducing a free closed 
string to an open string on the $Dp$-brane by imposing an orbifold 
condition:
$X^i (\s) = X^i (-\s)$, $X^a (\s) = - X^a (-\s)$,
$P^i (\s) = P^i (-\s)$, $P^a (\s) = - P^a (-\s)$.
If these constraints are imposed, the Hamiltonian is read as 
\beq
H &=& \half g_{ij} (\2pap) p^i p^j + 
\half \sum_{n=1} g_{ij} \left\{ (\2pap) K^i_n K^j_n +
\frac{n^2}{(\2pap)} Y^i_n Y^j_n \right\} \nn\\
& & + \half \sum_{n=1} g_{ab} \left\{ (\2pap) \bar{K}^a_n \bar{K}^b_n +
\frac{n^2}{(\2pap)} \bar{Y}^a_n \bar{Y}^b_n \right\}.
\eeq
Now the field theoretical propagator follows from integrating over 
the proper-time 
\beq \label{propa}
G[X_f; X_i] &=& \int^\infty_0 dT \langle X_f | e^{-iT \hat{H}} 
| X_i \rangle \\
%&=& \langle X_f | \frac{1}{H} | X_i \rangle \\
&=& i \int D[\Phi] \Phi[X_f] \Phi[X_i] \exp\left( -i \int D[X] 
\Phi[X] {\cal K} \Phi[X] \right) \nn
\eeq
where $\Phi[X] = \Phi[x^i, Y^i, \bar{Y}^a]$ and ${\cal K} =
\hat{H}$. Hence, the Hamiltonian in 
the first quantized theory corresponds to the kinetic operator for 
string field in the second quantized theory. 
%If we take the ghost sector into account the string field action 
%assumes the BRST invariant form
%\beq
%S_{SFT} = \int D[X, b, c] \Phi \{Q, b_0\} \Phi.
%\eeq

\section{Background of Tachyon Condensation}

The background of the tachyon condensation alters the boundary 
conditions for the open string on the D-brane. In order to have 
consistent equations of motion from the action Eq.(\ref{action})
we need to impose the following boundary conditions on $\p M$
\begin{mathletters} 
\beq
\left(-\frac{1}{\2pap} g_{ij} \p_\s X^j + 2 u_{ij} X^j 
\right)\Bigl\vert_{\s = \pi} = 0, \\
\left(\frac{1}{\2pap} g_{ij} \p_\s X^j + 2 u_{ij} X^j 
\right)\Bigl\vert_{\s = 0} = 0.
\eeq
\end{mathletters}
If we rewrite these boundary conditions in terms of normal modes,
we get
\begin{mathletters}
\label{bound:all}
\beq
\sum_n n X^i_n +i (\2pap) 2(g^{-1}u)^i{}_j \sum_n X^j_n &=& 0, 
\label{bound:a} \\
\sum_n n X^i_n (-1)^n -i (\2pap) 2(g^{-1}u)^i{}_j \sum_n X^j_n 
(-1)^n &=& 0. \label{bound:b}
\eeq
\end{mathletters}
In the framework of the canonical quantization we treat them as primary 
constraints. 
%Since the tachyon condensation affects the string 
%dynamics in the longitudinal directions only we are concerned with the 
%Hamiltonian $H_L$ which governs the string dynamics in the 
%longitudinal directions
%\beq
%H_L &=& \frac{s}{2} \sum_n g_{ij} \left((\2pap)P^i_{n} P^j_{-n} +
%\frac{n^2}{(\2pap)} X^i_n X^j_{-n} \right).
%-is u_{ij} \left(X^i X^j|_{\s =0} + X^i X^j|_{\s =\pi} \right)
%\eeq
%where we take $\int^{2\pi}_0 \frac{d\s}{2\pi}$. 
Let us denote the first constraint Eq.(\ref{bound:a}) as a primary 
constraint $\Phi^i_0$
\beq
\Phi^i_0 = \sum_n \left(n I +i(\2pap)2 g^{-1}u \right)^i{}_j X^j_n = 0.
\eeq
%Then we take its commutator with the Hamiltonian.
% using the fundamental brackets 
%\beq
%\{x^i, p_j\} = \delta^i{}_j, \quad
%\{X^i_n, P_{jn}\} = \delta^i{}_j \delta_{nm}.
%\eeq
Then the commutator of the primary constraint with the Hamiltonian 
generates a secondary constraint $\Psi_{i0}$, which is conjugate
to the primary constraint $\Phi^i_0$
\beq
\Psi_{i0} = \sum_n \left(n I - 2i (\2pap) 
ug^{-1} \right)_i{}^j P_{jn}=0.
\eeq
%(One may be concerned with the contribution of the boundary 
%term to the Hamiltonian. If we make use of the primary 
%constraints Eq.(\ref{bound:all}), we may write the boundary term in 
%the Hamiltonian as a bulk term,
%\beq
%-i u_{ij} \left(X^i X^j |_{\s=0} + X^i X^j |_{\s=\pi} \right)
%&=& \frac{1}{4\pi\a^\prime} \left(X^i g_{ij} \p_\s X^j |_{\s =0}
%- X^i g_{ij} \p_\s X^j |_{\s =\pi} \right) \\
%&=& - \frac{1}{4\pi\a^\prime} \int^\pi_0 d\s \p_\s \left(X^i 
%g_{ij} \p_\s X^j \right). \nn
%\eeq
%However, it is vanishing as the domain of $\s$ is extended to 
%that of the closed string, $[0, 2\pi]$. Thus, we may ignore it, 
%once we take the boundary conditions into account as primary
%constraints.)

The Dirac procedure requires further $\{H, \Psi_{i0} \} = 0$ and
it generates yet another constraint $\Phi^i_1$.
%\beq
%\Phi^i_1 = \sum_n \left(n^3 I + 2i(\2pap) n^2 g^{-1}u \right)^i{}_j 
%X^j_n = 0.
%\eeq
We may continue this procedure until it does not generates additional 
new constraints. By repetition we obtain a complete set of constraints
\begin{mathletters} 
\label{comp:all}
\beq
\Phi^i_m &=& \sum_n \left(n^{2m+1} I + 2i(\2pap) n^{2m} 
g^{-1}u \right)^i{}_j X^j_n = 0, \label{comp:a}\\
\Psi_{im} &=& \sum_n \left(n^{2m+1} I - 2i(\2pap) n^{2m} ug^{-1} 
\right)_i{}^j P_{jn}=0, \label{comp:b}
\eeq
\end{mathletters}
where $m = 0, 1, 2, \dots$. All these constraints belong to the 
second class. We may apply the same procedure to the primary 
constraint Eq.(\ref{bound:b}), but we only get a set of constraints 
equivalent to the set we already have. Thus, they are redundant.
It is quite useful to rearrange these set of constraints. 
From the constraints Eq.(\ref{comp:a}) it follows that
\beq
\sum_{m=0} \Phi^i_m \frac{(i\s)^{2m}}{(2m)!}
%&=& \sum_n \sum_{m=0} \left(n I + 2i (\2pap) g^{-1}u \right)^i{}_j
%\frac{(in\s)^{2m}}{(2m)!} X^j_n \\
= \sum_n \left(n I + 2i(\2pap) g^{-1}u \right)^i{}_j
\cos n\s X^j_n = 0. \nn
\eeq
If we make use of the following simple algebra,
$\int^{2\pi}_0 \frac{d\s}{\pi} \cos n\s \cos m\s = 
\delta(n-m) + \delta(n+m)$,
we find that the set of constraints $\{\Phi^i_m = 0, \,\,\,
m= 0, 1, 2, \dots \}$ is equivalent to  
\beq
\left\{x^i = 0,\,\,\,{\bar Y}^i_m = \frac{2}{m}(\2pap)(g^{-1}u)^i{}_j 
Y^j_m,\,\,\, m=1, 2, \dots \right\}. \label{consta}
\eeq
By a similar algebra, we conclude that the set of the constraints
$\{{\bar K}_{im} = 0, \,\,\, m = 0, 1, 2, \dots\}$ is equivalent to
the following set of constraints
\beq
\left\{p^i = 0, \,\,\,{\bar K}_{im} = \frac{2}{m}(\2pap)(ug^{-1})_i{}^j 
K_{jm}, \,\,\, m=1, 2, \dots \right\}. \label{constb}
\eeq
If the tachyon condensation does not occur, $u =0$, the 
constraints reduce to $\{{\bar Y}^i_m = {\bar K}_{im} = 0,
\,\,\, m = 1, 2, \dots \}$, i.e., the open string is attached to a flat
$Dp$-brane. As one of the parameters of the tachyon profile, $u_{pp}$ is 
turned on and reaches the infrared fixed point, 
$u_{pp} \rightarrow \infty$, the constraints for the canonical variables
in the direction of $p$ turn into the Dirichlet constraints, 
$\{Y^p_m = K_{pm} = 0,\,\,\, m = 1, 2, \dots \}$. Therefore, 
we find that the open string is now attached to a $D(p-1)$-brane.
 
If we exploit the explicit solution of the constraints, we easily
see how the Hamiltonian is deformed as the condensation develops.
Let us suppose that we turn on some of the tachyon profile 
parameters. Then the part of the Hamiltonian, which governs 
the dynamics of the canonical variables in the directions where
the profile parameters are turned on, may be written as 
\beq
H &=& \frac{(\2pap)}{2} \sum_{n=1} \left( {\bar K}_n g^{-1} {\bar 
K}_n + \left(\frac{n}{2}\right)^2 \frac{1}{(\2pap)^2} 
{\bar K}_n u^{-1} g u^{-1} {\bar K}_n \right) \nn\\ 
& & + \half \frac{1}{(\2pap)} \sum_{n=1} n^2 \left(
{\bar Y}_n g {\bar Y}_n + \left(\frac{n}{2}\right)^2 
\frac{1}{(\2pap)^2} {\bar Y}_n gu^{-1} g u^{-1}g{\bar Y}_n \right)
\eeq
Thus, as $u \rightarrow \infty$, 
it becomes the kinetic term for the open string variables along 
the Dirichlet directions
\beq
H &=& \frac{(\2pap)}{2} \sum_{n=1} {\bar K}_n g^{-1} {\bar K}_n
+ \half \frac{1}{(\2pap)} \sum_{n=1} n^2 {\bar Y}_n g {\bar Y}_n.
\eeq

\section{Open String Field Theory}

As the parameter of the tachyon profile is turned on, 
the string field and the Hamiltonian, equivalently the kinetic
operator in string field theory are deformed as we expect.
Now let us examine what effect the tachyon condensation background
makes on the string field action. Evaluating the Polyakov 
path-integral over a strip we obtain the kinetic part of the string 
field action
\beq
S = T_p \int D[X] \half \Phi[X] {\cal K} \Phi[X]
\eeq
where $T_p$ is the tension of the $Dp$-brane.
Let us suppose that we turn on only one of the tachyon profile 
parameters $u_{pp} = u$. Then, taking the constraints 
Eqs.(\ref{consta},\ref{constb}) into account, we may write the measure $D[X]$ as
\beq
D[X] &=& D[X^p] \prod_{i= 0, \dots, p-1} D[Y^i] \prod_{a=p+1,
\dots, d-1} D[{\bar Y}^a], \\
D[X^p] &=& d x^p \sqrt{g} \prod_{n=1} dY^p_n d{\bar Y}^p_n 
\delta \left({\bar Y}^p_n - \frac{2}{n} (\2pap) g^{-1} u Y^p_n \right)\nn
\eeq
where $g = g_{pp}$ and the metric $g_{ij}$ is diagonal.
If the tachyon condensation does not occur,
$\int D[X^p] \rightarrow \int D[Y^p]$.
As the system reaches the infrared fixed point of the condensation, 
where $u \rightarrow \infty$, the measure for the canonical variables
in the $p$-th direction becomes
\beq
\int D[X^p] &\rightarrow& \int d x^p \sqrt{g} \prod_{n=1} d{\bar Y}^p_n
\left(\frac{n}{\2pap}\right)\left(\frac{g}{2u}\right) \nn \\
& = & 2\pi\sqrt{\a^\prime} \sqrt{\frac{2u}{g}}
\int d x^p \sqrt{g} \prod_{n=1} d{\bar Y}^p_n
\eeq
where we make use of the zeta function regularization.

As the tachyon condensation is turned on the string becomes inactive
in the corresponding direction. Hence, it is appropriate to integrate
out $x^p$. In order to find the dependence of the string field 
$\Phi$ on $x^p$ we should be careful in defining the string 
propagator Eq.(\ref{propa}). Since the tachyon background term also 
contributes to the path-integral through the space-like boundary, 
the string propagator may be written as 
\beq
G[X_f;X_i] = \int^\infty_0 d T \langle X_f | e^{-iT \hat{H}}
|X_i \rangle e^{-\pi u x^2_p(T) - \pi u x^2_p(0)}.
\eeq
This expression is consistent with the analysis of the disk 
diagram in the boundary state formulation \cite{tlee}, 
which is related to the
open string field theory by the open-closed string duality.
Thanks to the constraints the propagator depends only on the zero
mode $x$ through the boundary action on the space-like boundary 
$e^{-\pi u x^2_p(T) - \pi u x^2_p(0)}$. It implies that the string
field $\Phi[X]$ can be factorized as 
\beq
\Phi[X] = e^{-\pi u x^2_p} \Phi[Y,\barY]
\eeq
in the infrared fixed limit. Hence, the string field action 
becomes in the infrared fixed limit
\beq \label{sftact}
S &=& 2\pi\sqrt{\a^\prime} T_p \sqrt{2u} \int d x 
e^{-2\pi u x^2} \int D[Y,\barY] \Phi[Y,\barY] 
{\cal K} \Phi[Y,\barY] \nn \\
&=& 2\pi\sqrt{\a^\prime} T_p \int D[Y,\barY] \Phi[Y,\barY] {\cal K} 
\Phi[Y,\barY]
\eeq
where $D[Y,{\bar Y}] = \prod_{i=0,\dots,p-1} D[Y^i]
\prod_{a=p,\dots,d-1} D[{\bar Y}^a]$.
Therefore, we find that the string field action for the open string
on a $Dp$-brane turns into that for the open string on a $D(p-1)$-brane
as the tachyon condensation develops. We also confirm the well-known relationship
between the D-brane tensions from Eq.(\ref{sftact})
$ T_{p-1} = 2\pi \sqrt{\a^\prime} T_p$.
Defining the string propagator we may construct the interacting 
open string field theory by gluing strings together. After the 
tachyon condensation occurs, the strings may be glued as usual.
Depending how the strings are glued, we get the Witten's 
cubic open string \cite{cubic}
%\beq
%S = 2\pi\sqrt{\a^\prime}T_p \int \left(\half \Phi {\cal K} \Phi + 
%\frac{1}{3} g \Phi \Phi \Phi \right)
%\eeq
or the covariant string field theory \cite{cova}.

\section{Conclusions}

We conclude this paper with a few remarks. As the tachyon condensation
develops, the $Dp$-brane turns into a lower dimensional D-brane. It 
has been well depicted in the boundary state formulation. Since 
the open string field theory and the boundary state formulation
are related by the open-closed duality, it is reasonable to expect
that the descendent transmutation of the D-branes may well be 
described in the framework of the open string field theory.
We find that the tachyon condensation background enters into 
the open string field theory through the constraints to be imposed
on the string variables. As the tachyon profile parameter varies,
the string field on the $Dp$-brane transmutes into that on a
lower dimensional D-brane. In this transmutation process it is also 
pointed out that the measure plays an important role in determining the 
tension of the lower dimensional D-brane. 

The effect of the tachyon condensation on the open string may be
seen more clearly as we evaluate the distance between two ends of 
the open string on the D-brane, which is given by
\beq
|X(0) - X(\pi)|^2 = \sum_{n,m=1} \frac{2(2n-1)(2m-1)}{(\2pap)^2}
\barY^i_n (gu^{-1}gu^{-1}g)_{ij} \barY^j_m.
\eeq
Since it is order of $1/u^2$, it vanishes in the infrared fixed 
limit. If the tachyon condensation takes place in every direction on 
the world-surface of the D-brane, the two ends of the open string
approaches to each other. Eventually in the infrared fixed point
limit they coincide and the open string turns into a closed 
string.

In the present work we discuss the canonical quantization of the 
open string in the background of the tachyon condensation. 
Although some important aspects of the tachyon condensation can 
be understood directly in the open string field theory, 
there remains much room for improvement to explore the full 
dynamical aspects of the tachyon condensation, including the
quantum corrections \cite{loop}.

\section*{Acknowledgement}
This work was supported by grant No. 2000-2-11100-002-5 from the Basic 
Research Program of the Korea Science \& Engineering Foundation. 
Part of this work was done during the author's visit to APCTP (Korea)
and KIAS (Korea).


\begin{thebibliography}{99}
\bibitem{com1} Note that in the boundary state formulation we take 
$S_T = i \int d\s T(X)$. The boundary state formulation is 
related to the canonical formulation of the open string theory 
under discussion by the double Wick rotation.
\end{thebibliography}
\end{document}